\documentstyle[rotate,times,epsfig]{l-aa}
\newif\iffigures

          \figurestrue 
\begin{document}
\thesaurus{03(11.14.1;11.05.2;11.06.1;11.19.3;11.05.1)}
\title{ A possible mechanism for the  mass ratio limitation in early type
galaxies}
\author{Yiping Wang \inst{1,2,3} \and Peter L.\ Biermann \inst{1,4}}

\offprints{ypwang@mpifr-bonn.mpg.de}

\institute{Max-Planck-Institut f\"ur Radioastronomie, Auf dem H\"ugel 69,
   D-53121 Bonn, Germany \and Purple Mountain Observatory, Academia
  Sinica, 210008 Nanjing, China \and ypwang@mpifr-bonn.mpg.de 
  \and plbiermann@mpifr-bonn.mpg.de}

\date{Received date Aug.11, 1997; accepted date}
\maketitle\markboth{Wang \& Biermann: A possible mechanism for the mass 
ratio limitation in early type galaxies}{Wang \& Biermann: A possible 
mechanism for the mass ratio limitation in early type galaxies}

\begin{abstract}
There is an interesting correlation between the central
objects and their host galaxies in recent high resolution HST
photometry of early type galaxies and Near-IR images of nearby quasar
hosts.  It has been shown that a) the hosts of these very luminous quasars 
are likely to be early type galaxies and that b) the mass ratio of
central black holes (BHs) and their host spheroidal components
($M_{\rm bh}/M_{\rm spheroid}$) is $\sim 0.002$
within a factor of three. Using the hierarchical galaxy evolution
scheme for the formation of early-type galaxies, we present here a
general viscous accretion disk model to trace the star formation and
central engine evolution before and after mergers. In our model,
starbursts and AGN coexist; these two activities compete for the
gas supply, interact with each other, probably feed back on each other 
and lock into a final status.  They thus constrain the ratio of central 
black hole mass and its host spheroidal mass to a universal value of order
$10^{-3}$.

\end{abstract}
\section{Introduction}
The main discussion about the formation of early-type galaxies centers
on a 'nature' versus 'nurture' dichotomy.  Sandage, Freeman \& Stokes
(1970) explained the 'nature' scheme as follows: A disk or a spheroidal
configuration is produced by the collapse of a rapidly rotating or
slowly rotating primordial gas cloud. The 'nurture' hypothesis, also
referred to as the hierarchical scheme (Toomre 1977, Kauffmann 1996,
Baugh et al. 1996, Mihos \& Hernquist 1996, Walker et al. 1996) is:
Mergers between disk galaxies where their mass ratio
($M_{companion}/M_{host}$) exceeds a value (of order $\sim ~0.3$
in some simulations, but almost certainly a function of orbital parameters) form
ellipticals or bulges in spiral galaxies; starbursts and nuclear
activity can be triggered probably by the merger event
(Fritze-v. Alvensleben \& Gerhard 1994).

It seems that the 'nurture' scheme, which models the observed
properties of ellipticals/bulges in a scenario where they are formed
by the merging of disk galaxies in a hierarchical universe, is more
supported by recent numerical simulations and observations (Barnes
1988, Kauffman et al. 1993), and it also, for the first time, placed
the theory of the formation and
evolution of elliptical galaxies in a proper cosmological context.

Massive compact objects appear to be ubiquitous components of galactic
nuclei.  Particularly attractive is the possibility that we are
observing the black holes (BHs) that once powered quasars and that still
provide the energy source of Active Galactic Nuclei (AGN).  This hypothesis is
at least crudely consistent with the observed properties of quasars and AGN.
Current black hole candidates have masses ranging from $\sim
10^{6.3}M_{\odot}$ in the case of Milky Way to $10^{9.5}M_{\odot}$ in
the case of M87 (Kormendy \& Richstone 1995). The compelling AGN
theory is that these massive BHs are fuelled by gaseous accretion disks
(L{\"u}st 1952, Lynden-Bell 1969, Shakura \& Sunyaev 1973, Rees 1984).

Starbursts and AGN are two fascinating phenomena after a merger. They
are usually treated as independent activities in numerical
simulations (Sargent et al. 1991, Barnes et al. 1992). Recent 
observations of Ultraluminous Infrared Galaxies (ULIGs) by Very Long
Baseline Interferometry (VLBI), the Very Large Array (VLA), infrared 
and optical wavelengths start to allow to consider a comprehensive picture 
for a starburst-AGN coexistence scenario (Smith et al. 1997, Heckman 1997). 
Also, HST photometry of ellipticals and spiral bulges, and Near-IR images 
of quasar hosts shows us that the central properties strongly depend on 
their host galaxies.  A 'nuclear luminosity/host-mass limitation' in the
most luminous quasars probably represents a physical limit on the 
mass of a black hole that can exist in a given galaxy spheroid mass. 
A mass ratio constraint of black hole mass and its host spheroid mass 
$\sim 0.002$ within a factor of three has been deduced from these observations 
and dynamical models (Faber et al. 1997, McLeod \& Rieke 1994a,1994b,
1995a,1995b, Magorrian et al. 1996, Kormendy \& Richstone 1995).  
The interesting point is whether this correlation is due to a statistical
sampling bias, or a substantial evolutionary result.

In this paper, we try to use a simplified model to describe the
steps of the elliptical/bulge formation, starburst and the
central engine evolution in a hierarchical scheme by assuming that
star formation and central object evolution are coexistent, and fed by a
viscous accretion disk. In our scheme a disk galaxy is entirely modelled
as an accretion disk.  Mergers can trigger a starburst in the central
region, and induce more turbulence there.  They thus drive turbulent
accretion to feed both the AGN and starburst, grow a massive BH (Linden et al. 
1993, Tacconi et al. 1994).  The competition and feedback interaction between
these two activities can lead to a constraint, and limit the ratio of BH mass
and its spheroidal stellar mass ($M_{\rm BH}/M_{\rm spheroid}$) to a certain
range of values.

\section{Model}
\subsection{Formation of early-type galaxies, star formation and
black hole evolution in a hierarchical universe}

\vspace{0.05cm}

We consider a hierarchical universe. In this scenario, an
elliptical/bulge does not form in a single collapse and burst of star
formation at high redshift. Instead, in our model, gas cools and
condenses at the center of a virialized halo of dark matter, and forms
a centrifugally-supported gas disk there; star formation takes place
mainly in the disk at a modest rate and gas is accreted inwards to the
central object at the same time; a stellar disk and a central black
hole may finally develop. If two disk galaxies merge at some time, the
violent interaction between the two galaxies will smash the original
stellar disks completely, and form the spheroidal component of the new early
type galaxy. Meanwhile, the activity drives the outer cold gas to the
central region of the newly formed galaxy and sets up a new molecular disk
there; the triggered starburst and central AGN will put their kinetic
energy into the ISM and induce turbulent viscous accretion in the
central region.

In the following, we describe the background for our model, and then
the model itself.

1) In the early universe, tidal torques spin up both the dark matter
and the baryonic matter in some region of space; when the region
becomes overdense, it will eventually collapse into a galaxy. This
process together with the cooling can finally form a primordial gas
disk, possibly with a seed black hole in the center, surrounded by a
dark halo.

2) When this primordial gas disk is formed, the viscosity can start to
transfer angular momentum to the outside and accrete material into the
central region of a galaxy, while at the same time, star formation is
taking place in the disk. The two activities, accreting material
inwards i)  to feed the central object and ii) to form a stellar disk, occur
together and outline a picture of evolution of the disk galaxy.

3) With the hierarchical galaxy formation model, the violent merger of
two disk galaxies will destroy the original stellar disks
completely. All the stars in the disks are transferred to the bulge or
spheroidal component of the new galaxy (Baugh 1996, Kauffmann et
al. 1993, Kauffmann 1995b, 1996), and thus form ellipticals/bulges.

4) During or after the violent mergers, the fate of the gas component
is very uncertain. Some gas may be shock-heated by collisions between
galaxies and injected back into the intergalactic
medium. Alternatively, a large fraction of the gas may lose its
orbital angular momentum, decrease its orbital radius, and be driven
to the center (Toomre \& Toomre, 1972). We propose here that most of
the gas contained within the central kiloparsec or so is in the form
of dense clouds which are on more or less circular orbits;
self-gravity may play a critical role in the subsequent evolution of
this molecular gas disk.

5) Both the concentration of molecular clouds towards the merger
nucleus and an increased efficiency of star formation due to
cloud-cloud collisions (Scoville et al. 1986) will result in the
appearance of a nuclear 'starburst'. This merger triggered starburst
probably can stir up turbulent accretion in the new molecular disk
by heating or shocking the ISM from supernovae (SN), and thus help to grow a quasar
black hole in a very short time. The possible fate of cool gas could
be to form stars in a starburst, to feed a quasar black
hole by turbulent accretion, or to be blown out of the disk by
a wind.

There are several key concepts and basic assumptions in our model:

1) By the disk formation scenario in the early universe (Dalcanton et
al.  1997), we basically assume that gas and baryons from a
protogalactic halo cool, collapse or settle into a rapidly rotating
gas disk on a time scale, that is much shorter than the star formation
time scale $t_{\ast}$ and the angular momentum transport time scale
due to spiral density waves etc. (Zhang 1996, Gnedin et al. 1995,
Olivier et al.  1991).  Nevertheless, this assumption does not
restrict our results here. Because, even if there is still some gas in
a hot phase when the merger takes place, this gas could cool and
together with the cool molecular gas from the outer disk set up the
new molecular disk in the central region of the elliptical. This can
change the total mass of the newly formed molecular disk, increase
both star formation and accretion to the central black hole, but it
will not affect the final mass ratio of the black hole and its stellar
component.

2) We basically assume that the first merger event between two disk galaxies
which possibly form the elliptical/bulge happens around $10^{9} yr$ or 
later after the gas and baryons from a protogalactic halo cool, collapse 
and initially set up a gaseous disk. At this later evolution stage,
the drastic disk evolution has already quieted down, or in
other words, a stellar disk has already been well developed (Rieke
et al. 1980, Kronberg et al. 1985). This merger time scale fits also the
cosmological pattern of the merger event and the formation history of
ellipticals/bulges near $\rm z\approx 1$ (Baugh et al. 1996), but we can see
in Fig.5 that the final mass ratio limitation does not depend on the
exact merger time.

3) With the basic assumption that $t_{\ast}$ is proportional to
$t_{\rm acc}$ (i.e. $t_{\ast}= \alpha t_{\rm acc}, \alpha \simeq 1$), Lin \& Pringle
(1987) got a nice fit for the exponential stellar disk evolution in a
spiral galaxy.  We will adopt the same assumption for our disk
evolution and also for the starburst and AGN stage after the merger. We
found that the final mass ratio limitation $\sim 0.006$ strongly
depends on this assumption of equal time scales (see Fig.10).

4) As for the uncertainty of the detailed physical interaction between
starburst and AGN, the purpose of this paper is just to
sketch out an indicative picture for these activities; we will
basically imitate the coexistent `starburst' and the subsequent
AGN evolution with the starburst time scale and the
turbulent accretion time scale much shorter than their normal time
scales before mergers (Mihos \& Hernquist 1994).

5) We adopt the notion that accretion onto the central BH is
restricted to a rate at which the Eddington luminosity is reached. In our
calculation, we assume that the central BH is a seed BH ($\sim
5M_{\odot}$) at the outset, the growth of this seed BH is limited by
the Eddington limit. The extra gas supply to the center can form
gas clouds there shrouding the central BH, or be blown away by a wind.

According to the scenario sketched here, disk evolution including the
star formation in the gas disk and a viscous accretion to the central
region will start in a protogalaxy before the merger, thus develop a
stellar disk and a black hole in the center. We will show the BH
evolution, stellar mass evolution and mass ratio evolution for a
normal disk galaxy in Fig.1 and Fig.2 (line).  By our model, we
assume, at a certain time, after the disk evolution quiets down, two
disk galaxies merge, firstly, destroy the original stellar disks and
form the spheroidal component, secondly, drive the cool gas left to
the center, redistribute a molecular disk there. A starburst can be
triggered in this newly formed molecular disk by the violent
interaction, and it probably will induce a turbulent viscous accretion
to feed both the central AGN and starburst there. These two activities 
will grow
together rapidly, compete for the gas supply and drain the gas in
the molecular disk in a very short time, thus restrict the growth of
the black hole to a certain ratio relative to the newly formed stellar
component. We will show that the mass ratio of these two components
($M_{\it bh}/M_ {\it spheroid}$) converges to a limited range of value $\sim
0.006$ from our calculation in Fig.2 (dashed line).

\subsection{ Fundamental equations and viscosity}

\vspace{0.05cm}

We take the gas surface density $\Sigma$, to be the sum of HI and
H$_{\rm 2}$ surface density (plus the small amount of heavy elements).
The evolution of $\Sigma(R,t)$, is governed by the standard viscous
accretion disk equations of continuity and conservation of angular
momentum (L{\"u}st 1952, Pringle 1981), with a sink term due to the
conversion of gas to stars ($\Sigma/t_{\ast}$). The basic equations are

\begin{equation}
\label{eq.masscon}
\frac{\partial \Sigma}{\partial t} + \frac{1}{R}
\frac{\partial}{\partial R}\left(\Sigma R V_{R}\right) = -
\frac{\Sigma}{t_{\ast}}\left(1-R_{\rm e} \right)
\end{equation}

\begin{eqnarray}
\frac{\partial}{\partial t}\left(\Sigma R^2 \Omega\right) +
\frac{1}{R}\frac{\partial}{\partial R}\left(\Sigma R^3 \Omega
V_{R}\right) &=&
\frac{1}{R}\frac{\partial}{\partial R}\left(\nu \Sigma R^3
\frac{\partial \Omega}{\partial R} \right)\nonumber\\
&&-\frac{R^2 \Omega \Sigma}{t_{\ast}}\left(1-R_{\rm e} \right)
\end{eqnarray}

If we assume $\dot\Omega=0$ for a stable accretion, and replace
the radial velocity $V_{\rm R}$ from equation (1) and (2), we get,

\begin{eqnarray}
\frac{\partial \Sigma}{\partial t} =
-\frac{1}{R}\frac{\partial}{\partial R} \left\{
\left[ \frac{\partial (R^2
\Omega)}{\partial R} \right]^{-1}
\frac{\partial}{\partial R} \left( \nu
\Sigma R^3 \frac{\partial \Omega}{\partial R}\right) \right\}\nonumber\\
 -\frac{\Sigma}{t_{\ast}}\left(1-R_{\rm e} \right)
\end{eqnarray}

Where $R_{\rm e}$ is the mass return of gas through mass loss from 
stars.  In our calculation, we take $R_{\rm e} \cong 0.3$ (Tinsley
1974), but we show in Fig.6 how sensitive the mass ratio limitation is
to this fraction. $\Omega$ is the angular velocity, and $t_{\ast}$ is
the star formation time scale. As for the viscosity $\nu$, we will
take a new viscosity prescription $\nu =\beta_{1} v_{\rm \phi} r$
(Duschl et al. 1997) for our Keplerian selfgravitating molecular
disk. In this case, the accretion time scale $t_{\rm acc} = r^2/\nu =
r/\beta_{1}v_{\rm \phi} $ (Pringle 1981).

In our model, we basically assume that $t_{\ast}$ is proportional to
$t_{\rm acc}$, i.e. $t_{\ast}=\alpha t_{\rm acc} = \beta_{2} r/v_{\rm
\phi}$ for the disk evolution stage before merger and also the
starburst \& AGN stage after a merger. In this case, with a flat rotation
law , we get the star formation rate as: $\frac{\partial
\Sigma_{\ast}}{\partial t} \propto \Sigma \Omega$, which is exactly
the observed star formation law for disk galaxies (Dopita et
al. 1994). We will show in Fig.10 that the relationship of these two
time scales $t_{\rm acc}/{t_\ast} = (\beta1\beta2)^{-1}$ is a key
parameter for the final mass ratio limitation.

The evolution of the stellar surface density ${\partial
\Sigma_{\ast}}/ {\partial t}$ and the central black hole $\partial
M_{\rm BH}/{\partial t}$ are given by the prescriptions as below:

\begin{equation}
\frac{\partial \Sigma_{\ast}}{\partial t}=\Sigma/t_{\ast}
\end{equation}

\begin{equation}
\frac{\partial M_{\rm bh}}{\partial
t} = -2\pi\beta_{\rm 1}V_{\rm 0}\left[ 2R\Sigma + R^{2}\frac{\partial
\Sigma}{\partial R}\right]_{R=R_{\rm in}}
\end{equation}

Where $R_{\rm in}$ is the inner boundary radius of the molecular disk.
We basically choose $R_{\rm in} \cong 0.1pc $ in our calculation, a 
value which we adopt for the inner edge of the molecular disk (Moran 
et al. 1995, Barvainis 1995). We will show in Fig.7 that the final
mass ratio limitation weakly depends on this radius. 
$V_{\rm 0} = 100 km/s$ is a standard velocity in our calculation, with
the relationship $v_{\rm \phi} \propto V_{\rm 0}$. 

The stars are assumed to be on approximately
circular orbits and to stay at the radii at which they are formed, so
they are effectively frozen out of the viscous evolution.  For that
gas left still in the disk, it will continue the viscous accretion
process.

As for most disk galaxies, their rotation curves are usually quite
flat outside the parsec scale, which corresponds well to the region of
our interests (Linden et al. 1993, Yoshiaki et al. 1989, de Blok et
al. 1996, Genzel et al. 1987, 1994). Our own galaxy shows rather well,
that a flat rotation is a fair approximation in the inner
region. Also, the recent HST work by Faber et al. (1997) shows that
even the general class of ``core galaxies'' with an inner surface
brightness law of a powerlaw in radius with an index 0. to -0.3
still give a rather shallow radius dependence of their circular
velocity with $v_{\phi} \sim r^{0.5 ..0.35}$, which corresponds to an
angular velocity law of $\Omega \sim r^{-0.5 .. -0.65}$, still far from rigid
rotation.  The other class of ``power
law galaxies'' all have an inner rotation law which is quite flat.
So, we adopt the approximation of a  flat rotation curve in our
calculation, and solve the equations (3), (4) and (5) by a standard
first-order explicit finite difference method.

\section{Calculations and discussion}

\subsection{ The numerical scheme and boundary conditions}

\vspace{0.05cm}

For computational convenience to solve the partial differential
equation (3), we introduce here dimensionless variables,
$r'=r/r_{\rm 0}$, $ t'=t/t_{\rm 0}$, $ \Sigma'=\Sigma/\Sigma_{\rm 0}$
with the scale $r_{\rm 0}=1 \, \rm pc$, $t_{\rm 0} \sim ~10^{4}yr$, the
general boundary conditions as $lim_{r \rightarrow
R_{out}}[\Sigma(r,t)]
\longrightarrow 0$, and zero torque at the origin, $lim_{r\rightarrow R_{in}}
[G(r,t)]=0$, where $R_{out}$ and $R_{in}$ are the disk's outer radius
and inner radius, and the viscous torque $G$ as given by (Pringle 1981).  

$$
G(r,t)=2 \pi \nu \Sigma R^{3} \frac{\partial \Omega}{\partial R} 
$$

This choice of boundary conditions guarantees zero viscous coupling
between the disk and the central object, and so allows all mass
reaching that point to flow freely inward.  Evidently, the accretion to the
central BH is still limited by the Eddington condition.

\subsection{ Initial conditions}

\vspace{0.05cm}

The procedure of our calculation is at first to develop a disk galaxy,
and after the merger, to continue the turbulent viscous accretion disk
evolution in the newly formed molecular disk.  For this purpose, we
set up two reasonable initial density distributions for the gas
disks before and after a merger. We will discuss these two initial
conditions in paragraph 1) and 2) below.

1) The initial gas disk surface density for disk galaxy formation is
dependent on the characteristics of the protogalactic halo and on the
details of dissipative collapse of the gas to the rotationally
supported disk. As for our disk evolution scenario, our disk is
basically a Keplerian selfgravitating disk (Duschl et al., 1997), we
will just use its stationary solution $\Sigma(R) \propto
\sqrt{R_{\rm 0}-R}/R$ as our initial gas disk surface density
distribution for the disk evolution, where $R_{0}$ is the outer radius
of our disk $\sim ~20 \, \rm kpc$.

2) In our calculation, we will assume the initial gas distribution of
the molecular disk assembled after merger to be a power law
$\Sigma(R)\propto 1/R$, as a good approximation to the equilibrium
distribution of mass with a highly flattened axisymmetric,
self-gravitating system (Toomre 1963).  This law fits the observation
result of the distribution of molecular gas in merger remnants, such
as, ARP220 (Scoville et al. 1986), where more than $70\%$ of the CO
emission, corresponding to more than $\sim 10^{10} \, M_{\rm \odot}$, is
confined to a galactocentric radius smaller than 1 kpc.

\subsection{ Numerical calculation and discussion }

\vspace{0.05cm}

In our model, we basically consider the case when a merger happens
some time after the disk has already been well developed. This does
not mean that it is a necessary condition for merger in real life, but
it does really exclude the possibility that ellipticals or bulges form
through a single collapse of a protogalaxy in high redshift.

In order to understand the merger effects, we first show in Fig.1 and
Fig.2 (line) a normal disk evolution. We can see in Fig.1 that after BH
evolution starts from a seed BH ($\sim 5 \, M_{\odot}$), it will grow a
central BH $\sim 10^{8} \, M_{\odot}$ (for a protogalaxy of
$10^{11}M_{\odot}$) after a full Eddington evolution of nearly
$10^{9}$ years. Because the gas was almost used up by the BH evolution and
the star formation finally, both activities will quiet down, thus
the ratio of the BH mass and the stellar mass will slowly approach a
constant. We emphasize that no spheroid exists at all during this disk
evolution stage.

According to our model, at a certain time after the disk evolution
quiets down, if a violent merger happens between two well developed
disk galaxies, it will smash the original stellar disks completely and
form the spheroid of a new early type galaxy. Meanwhile, the merger can
cause a severe perturbation in the central region and drive the gas
inwards to concentrate in the central kiloparsec (Barnes 1992, Casoli
et al. 1988). We simulate this effect by redistributing a certain amount
of cool gas to that region, setting up a molecular disk, triggering
a starburst and nuclear activity there. In the real universe, the
violence of the merger to form an elliptical/bulge can have a variety of
properties. So, the amount of gas redistributed in the center by
the merger effect can be different also.  In our simulation, we
therefore redistribute the same amount of cool gas as the host galaxy's in
its central region, to simulate the effect of a major merger between two
disk galaxies with comparable mass, and half the amount of cool gas for
the case of a minor merger. We find from our calculation that the amount
of gas mass redistributed to the central region has no influence on the
mass ratio limitation, but it does affect the final black hole mass.

\begin{figure}
\iffigures
\setlength{\unitlength}{1cm}
\begin{picture}(18,8.7)
\put(0.7,0.7){\psfig{figure=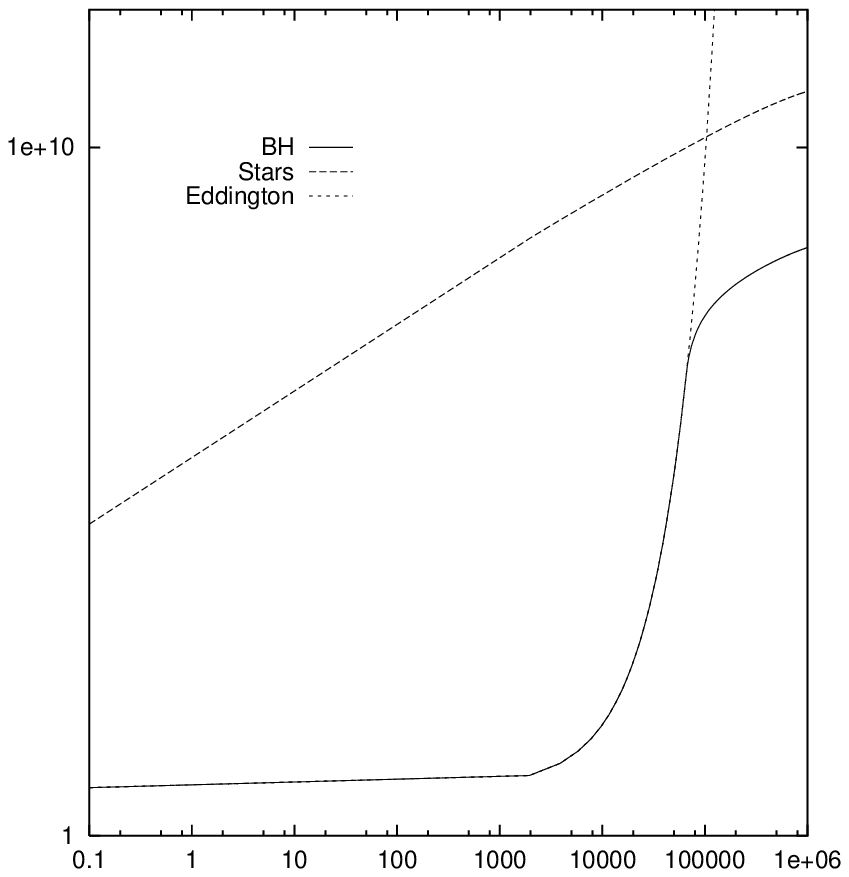,width=8 cm}}
\put(4.0,3.5){\Large \bf nonmerger}
\put(3.9,0.2){$(t/t_{\rm 0}), t_{\rm 0}=10^{4}yr$}
\put(0.5,3.9){\rotate{$Mass(M_{\odot})$}}
\end{picture}
\else
\picplace{6.0cm}
\fi
\caption[]{\label{nonmerger} The temporal evolution of the
central black hole mass and the stellar mass in a normal disk galaxy 
(without any merger). The short dashed curve is the Eddington limitation
for the growth of BH.  We note that BH growth at the Eddington limit in a
log-log plot of mass versus time is exponential.} 
\end{figure}
\begin{figure}
\iffigures
\setlength{\unitlength}{1cm}
\begin{picture}(18,8.7)
\put(0.7,0.7){\psfig{figure=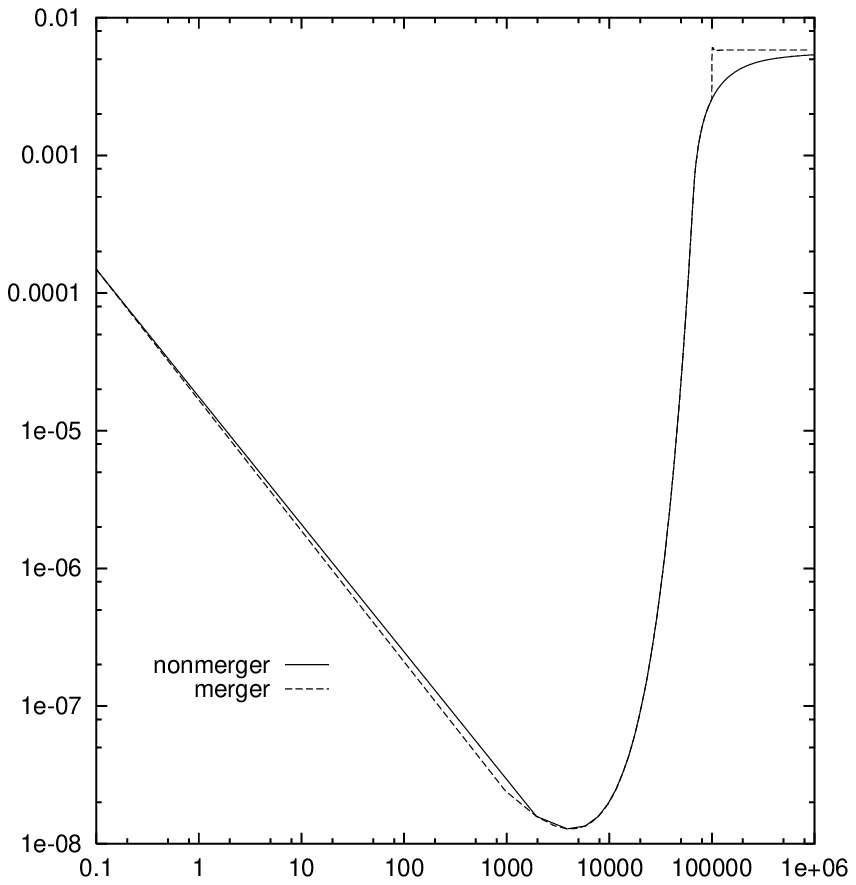,width=8 cm}}
\put(5.7,7.8){$merger \rightarrow$}
\put(3.9,0.2){$(t/t_{\rm 0}), t_{\rm 0}=10^{4}yr$}
\put(0.5,3.9){\rotate{$M_{\rm bh}/M_{\rm spheroid}$}}
\end{picture}
\else
\picplace{6.0cm}
\fi
\caption[]{\label{mass4} The time evolution of the mass ratio ($M_{\rm bh}/
M_{\rm spheroid}$) in a normal disk galaxy (without any merger), and that
with one merger.  We note that both cases lead to a very similar final mass
ratio.} 
\end{figure}

We show in Fig.2 (dashed line) and Fig.3 the evolution of the mass
ratio ($M_{\rm bh}/M_{\rm spheroid}$), BH mass and the stellar
component mass for the case when a major merger happens near
$10^{9}\rm yr$.  We can see that black hole grows in full Eddington
evolution at the beginning, until some time later, the gas in the disk
is drained off by star formation and BH evolution, both activities
quiet down, thus constrain the mass ratio firstly around a limited
value of $\sim 0.006$. If a merger happens afterwards, it can rewaken
these two activities, with the appearance of a starburst and central
AGN. The extreme starburst and BH growth last only for a short
time, then the gas in the central region will be used up by these two
activities, or some of it is blown away by the wind; till this stage,
we see the normal ellipticals/bulges with the mass ratio limited. From
Fig.2 and Fig.3, we see that a merger dramatically shortens the
convergence time scale and grows a massive BH in the center quickly.

In our model, the prescriptions of the star formation time scale and
the accretion time scale are $t_{\ast}= \beta_{2} r/v_{\rm \phi}$, and
$t_{\rm acc} = r/\beta_{1}v_{\rm \phi} $. Considering a reasonable
accretion time scale ($10^{9} \sim 10^{10}yr$) and the star formation
rate for a normal galaxy (Dalcanton et al. 1997), we choose
$\beta_{1}=0.005, \beta_{2}=200$ for the normal disk evolution phase.
We simulate the starburst and central engine activity by taking
$\beta_{1}=0.5, \beta_{2}=2$, thus shorten both time scales by a
factor of 100.  So it will dramatically increase the star formation
rate to nearly two orders of magnitude in ultraluminous starburst
galaxies; it also fits the starburst time scale about $10^{7}\sim
10^{8}yr$ (Rieke et al. 1980, Kronberg et al. 1985, Mihos \& Hernquist
1994, Smith et al.  1996). We will show in Fig.9 and Fig.10 that the
mass ratio is not very sensitive to the exact number of $\beta_{1}$
and $\beta_{2}$, but clearly sensitive to their correlation. We will
discuss this in detail below.

We show the mass ratio evolution in Fig.4, for the case that
protogalaxy mass ranges from $\sim 10^{11} \, M_{\rm \odot} $ to $\sim
10^{12} \, M_{\rm \odot}$; and in Fig.5, the case when merger happens at a
different time after the disk quiets down. We see from our results
that the final mass ratio limitation does not depend on these two
parameters.

In our calculation, we basically choose the star formation mass return
rate $\rm R_e = 0.3$ (Tinsley 1974), and the inner radius for the
molecular disk $R_{\rm in} = 0.1 \, pc$ (Moran et al. 1995, Barvainis
1995). In Fig.6 and Fig.7, we see that the mass ratio limitation will
vary moderately if we vary $R_{\rm e}$ by a factor of two and
$R_{\rm in}$ by a factor of 12.

From our calculation, we see that the mass ratio of the central black
hole and its host spheroid is very sensitive to the ratio of the
starburst time scale and the turbulent accretion time scale after a merger (see
Fig.10). The mass ratio converges to the value $\sim ~0.006$
only when the turbulent accretion time scale approximately equals the
starburst time scale (i.e. $\beta_{1} \ast \beta_{2}\cong$ 1).
This probably hints at a physical reality, which is, a merger will drive
a large amount of cold gas inwards to the central region, the
increased molecular cloud collisions can trigger a starburst in the
central region, the kinetic energy output from these young massive
stars and the shocks from supernovae will heat and disturb the ISM
locally, thus probably induce turbulent viscous accretion to feed both the
AGN and starburst in the central region. These two activities,
starburst and central AGN, will interact with each other, feedback,
drain the gas in the molecular disk in a short time, thus grow a massive
BH quickly there. To some degree, it probably can start a strong wind
from the center, blow off the leftover gas in the disk, stop the starburst
and the accretion process in the central region. It seems that this
evolution scenario can help to explain the observed limited mass ratio
region $\sim 0.002$ within a factor of three. In order to see the
sensitivity of this mass ratio limitation to the equal time scale
assumption, we show the results in Fig.10 for different ratios of the
starburst time scale and the turbulent accretion time scale;
we can see the dramatic divergence of the mass ratio when 
$t_{\rm acc}/t_{\ast} = 0.1, 1, 10$.

All we discussed above is the case where only one major merger happens
between two disk galaxies to form an elliptical or the bulge of a
spiral. In a real hierarchical universe, multiple minor mergers would
also probably occur after the major merger. These could bring in more
cool gas to the central region and start another evolution cycle. In
order to imitate our universe closely, we also calculated the case
that one major merger is followed by another minor merger. We show our
result in Fig.8, and it seems that the multimerger will not change
this universal ratio.

\begin{figure}
\iffigures
\setlength{\unitlength}{1cm}
\begin{picture}(18,8.7)
\put(0.7,0.7){\psfig{figure=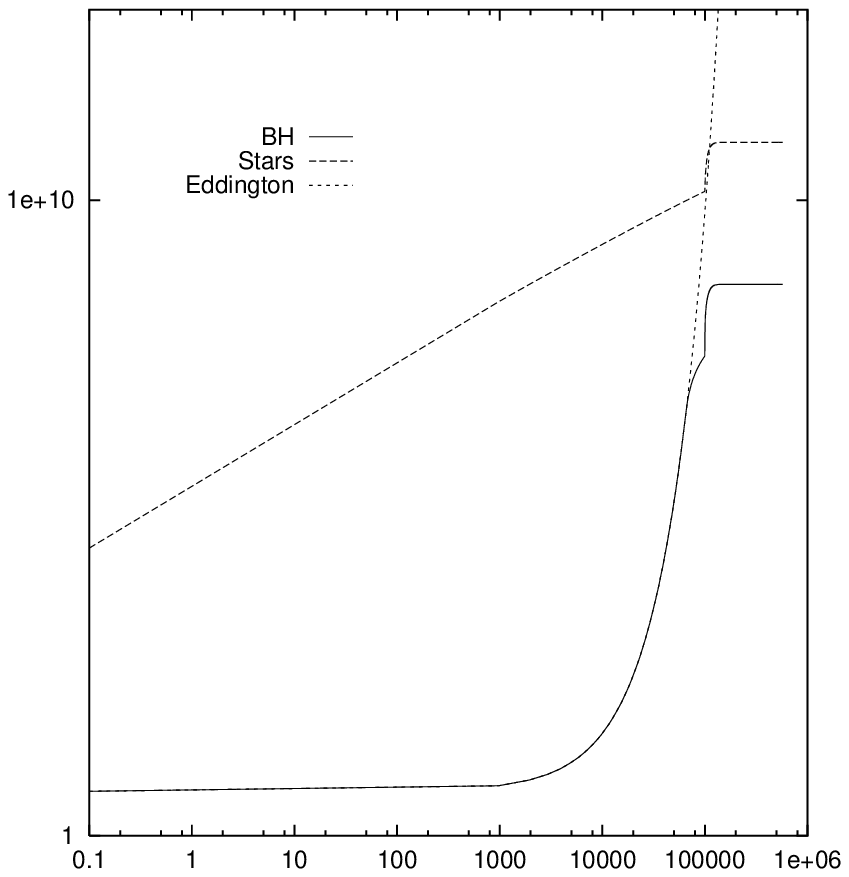,width=8 cm}}
\put(4.0,3.5){\Large \bf merger}
\put(6.9,5.4){$\rhd$}
\put(6.9,6.9){$\rhd$}
\put(4.0,0.2){$ (t/t_{\rm 0})$}
\put(0.5,3.7){\rotate{$Mass(M_{\odot})$}}
\end{picture}
\else
\picplace{6.0cm}
\fi
\caption[]{\label{mass1} The time evolution of the black hole mass,
spheroidal component mass and the Eddington growth mass limitation
with one major merger. The triangles indicate the merger event.}

\end{figure}

\begin{figure}
\iffigures
\label{mass}
\setlength{\unitlength}{1cm}
\begin{picture}(18,8.7)
\put(0.7,0.7){\psfig{figure=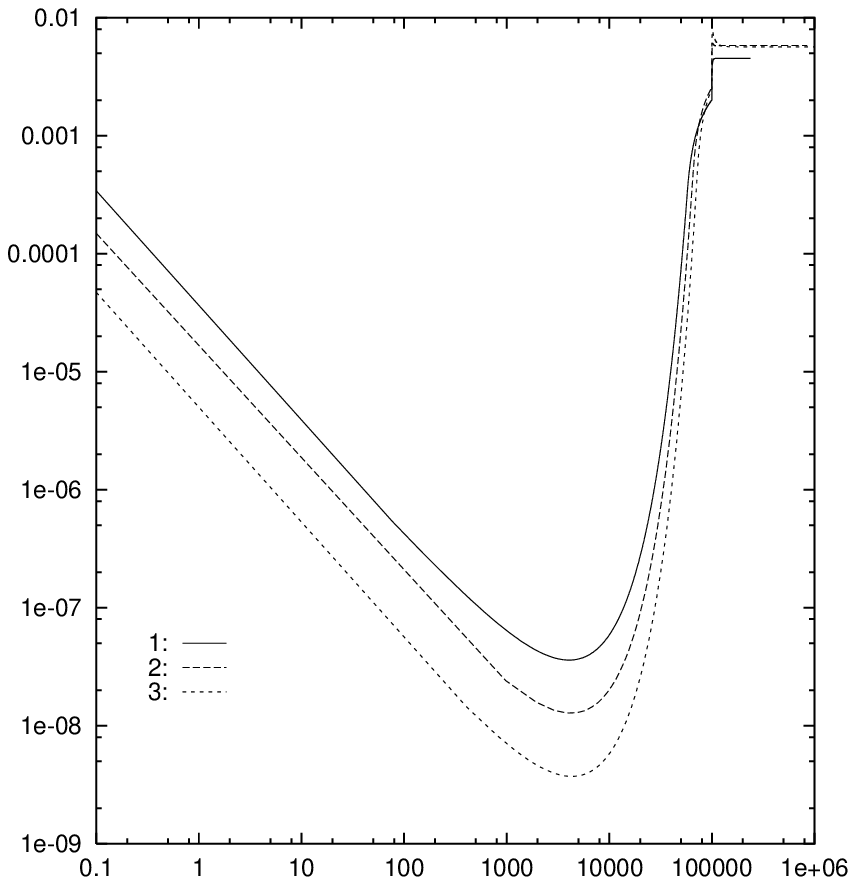,width=8 cm}}
\put(4.0,0.2){$ (t/t_{\rm 0})$}
\put(2.4,7.4){$protogalaxy$}
\put(2.5,7.0){$1: 10^{11}M_{\odot}$}
\put(2.5,6.6){$2: 10^{12}M_{\odot}$}
\put(5.7,7.8){$ merger \rightarrow $}
\put(0.5,3.7){\rotate{$M_{\rm bh}/M_{\rm spheroid}$}}
\end{picture}
\else
\picplace{6.0cm}
\fi
\caption[]{\label{mass} The time evolution of mass ratio of the central
black hole and its host spheroid for the protogalaxy mass ranges from
$10^{11}M_{\odot}$ to $10^{12}M_{\odot}$. The curves coincide after the merger, and
converge to the almost same mass ratio $\sim 0.006$.}
\end{figure}

\begin{figure}
\iffigures
\setlength{\unitlength}{1cm}
\begin{picture}(18,9.7)
\put(0.7,0.7){\psfig{figure=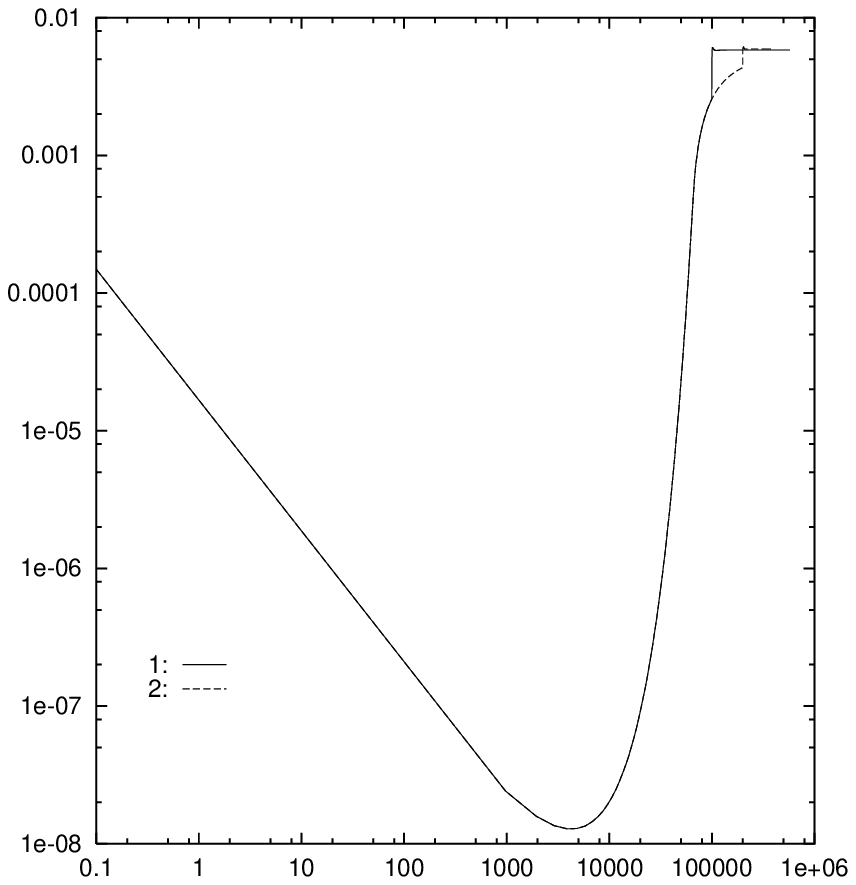,width=8 cm}}
\put(4.0,0.2){$ (t/t_{\rm 0})$}
\put(2.8,6.4){$1: t_{\rm merger}=10^{9}yr$}
\put(2.8,6.0){$2: t_{\rm merger}=2\times 10^{9}yr$}
\put(5.8,7.8){$ merger \rightarrow$}
\put(7.6,7.8){$\uparrow$}
\put(0.5,3.5){\rotate{$M_{\rm bh}/M_{\rm spheroid}$}}
\end{picture}
\else
\picplace{6.0cm}
\fi
\caption[]{\label{mass2} The time evolution of the mass ratio
($M_{\rm bh}/M_{\rm spheroid}$)
when the merger happens in different times.}

\end{figure}

\begin{figure}
\iffigures
\setlength{\unitlength}{1cm}
\begin{picture}(18,8.7)
\put(0.7,0.7){\psfig{figure=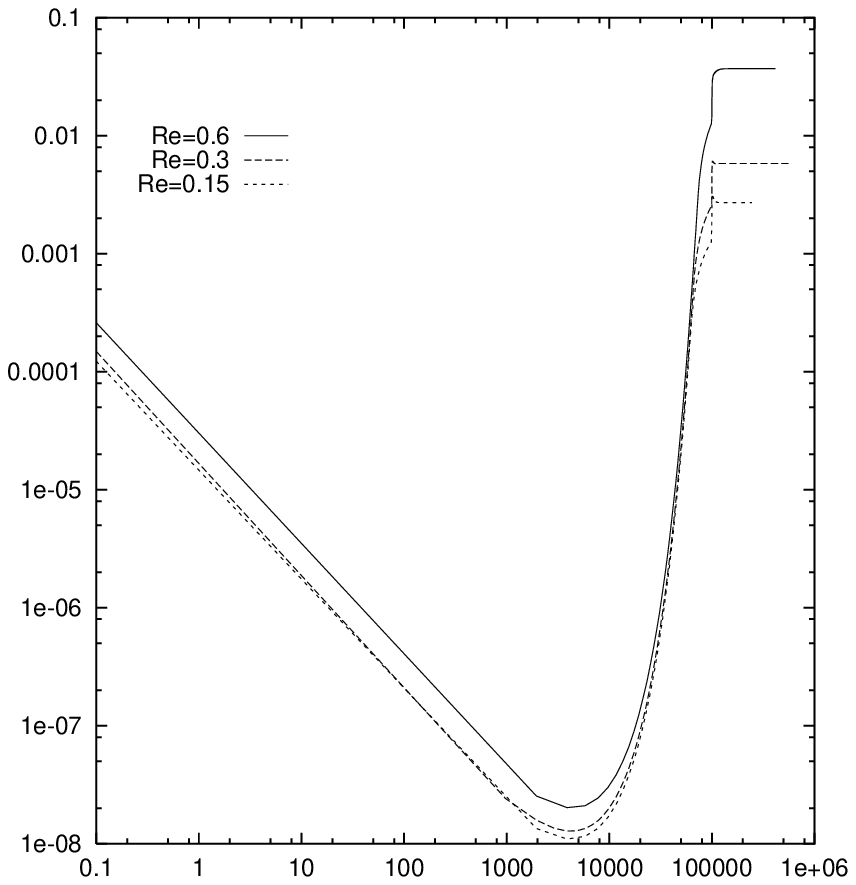,width=8 cm}}
\put(4.0,0.2){$ (t/t_{\rm 0})$}
\put(0.5,5.0){\rotate{$M_{\rm bh}/M_{\rm spheroid}$}}
\end{picture}
\else
\picplace{6.0cm}
\fi
\caption[]{\label{delay} The time evolution of the mass ratio of the
central black hole and its host spheroid for the mass return rate 
$R_{\rm e} =
0.6, 0.3, 0.15$.  A standard value for this parameter is 0.3 (Tinsley 1974),
but various common IMFs can give other values. Varying this parameter 
by a substantial factor does influence the resulting mass ratio $M_{\rm bh}/ 
M_{\rm spheroid}$. It appears as if this final ratio depends on 
$R_{\rm e}$
approximately as $(1-R_{\rm e})^{-3.4}$. This means that our results depend on 
the assumption that the universal spread of this parameter is less than 
about two.}

\end{figure}

\begin{figure}
\iffigures
\setlength{\unitlength}{1cm}
\begin{picture}(18,8.7)
\put(0.7,0.7){\psfig{figure=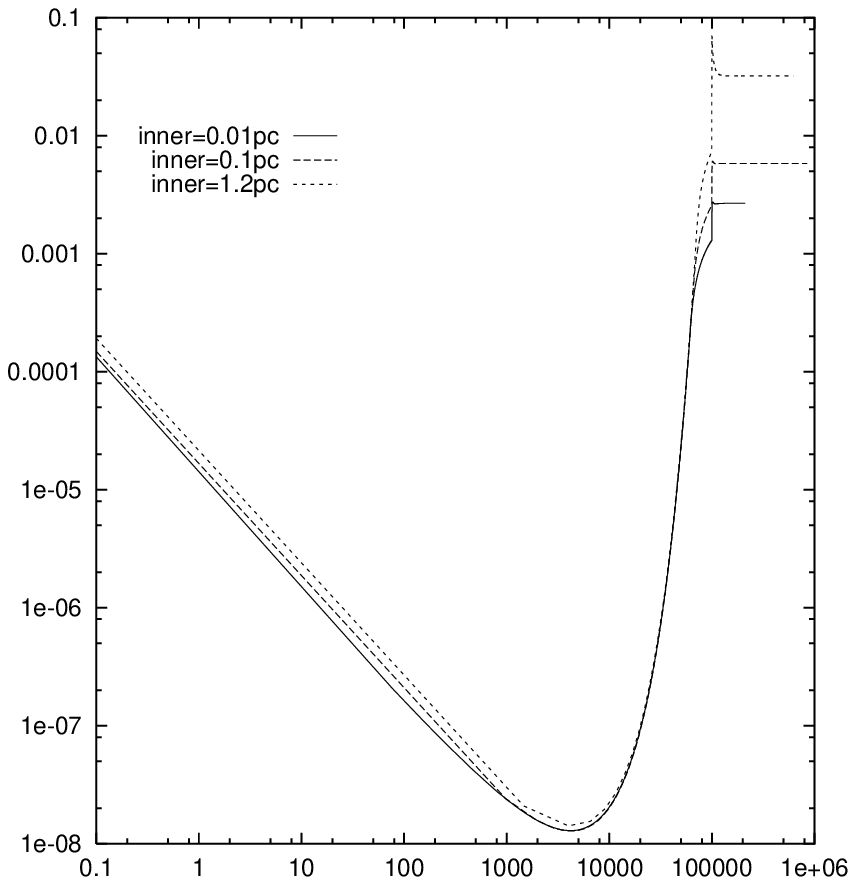,width=8 cm}}
\put(4.5,0.2){$ (t/t_{\rm 0})$}
\put(0.5,4.0){\rotate{$M_{\rm bh} (M_{\rm spheroid})$}}
\end{picture}
\else
\picplace{6.0cm}
\fi
\caption[]{\label{delay1} The time evolution of mass ratio of central
black hole and its host spheroid when the inner boundary radius of the
molecular disk shifts with a factor of 12, i.e. $R_{\rm in} = 0.1pc,
1.2pc$. We see in the graph that the mass ratio converges to a
constant quickly after the merger. The final ratio $M_{\rm bh}/
M_{\rm spheroid}$ varies by a factor of 5 for
the variation of $R_{\rm in}$ by a factor of 12. This shows a soft dependence.}

\end{figure}

\begin{figure}
\iffigures
\setlength{\unitlength}{1cm}
\begin{picture}(18,8.7)
\put(0.7,0.7){\psfig{figure=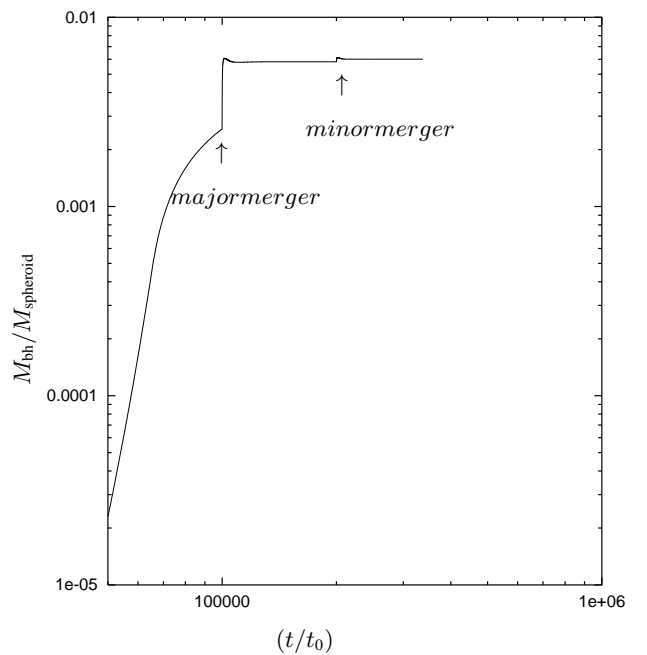,width=8 cm}}
\put(4.0,0.2){$ (t/t_{\rm 0})$}
\put(3.2,6.7){$\uparrow$}
\put(2.6,6.1){$ major merger$}
\put(4.8,7.6){$ \uparrow $}
\put(4.4,7.0){$ minor merger $}
\put(0.5,3.7){\rotate{$M_{\rm bh}/M_{\rm spheroid}$}}
\end{picture}
\else
\picplace{6.0cm}
\fi
\caption[]{\label{moremerger} The time evolution of the mass ratio of the
central black hole and its host spheroid with one major merger and another
minor merger.}

\end{figure}
\begin{figure}
\iffigures
\setlength{\unitlength}{1cm}
\begin{picture}(18,8.7)
\put(0.7,0.7){\psfig{figure=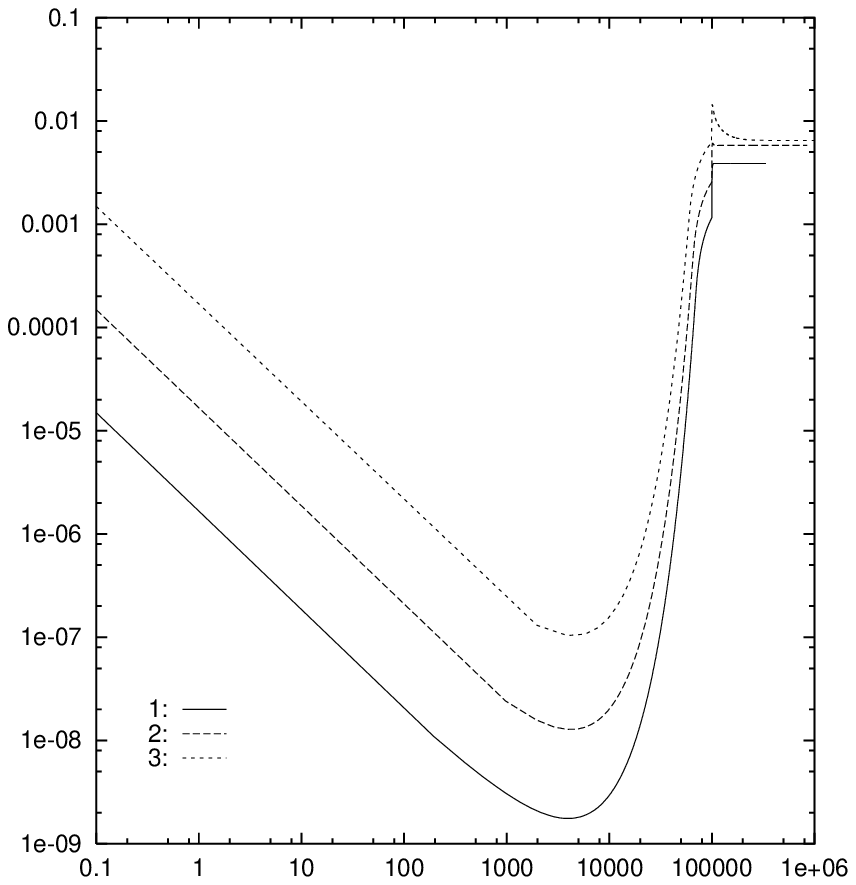,width=8 cm}}
\put(4.0,0.2){$ (t/t_{\rm 0})$}
\put(2.2,7.4){$1: \beta_{1}=0.05;\beta_{2}=20$}
\put(2.2,7.0){$2: \beta_{1}=0.005;\beta_{2}=200$}
\put(2.2,6.6){$3: \beta_{1}=0.0005;\beta_{2}=2000$}
\put(0.5,3.7){\rotate{$M_{\rm bh}/M_{\rm spheroid}$}}
\end{picture}
\else
\picplace{6.0cm}
\fi
\caption[]{\label{tscaledif} The time evolution of the mass ratio of the
central black hole and its host spheroid when the parameters $\beta_{1},
\beta_{2}$ vary with a factor of 10. We see in the graph that the
final mass ratio limitation does not strongly depend on the exact value of
these parameters, as long as $\beta_{1} \ast \beta_{2} \cong 1$.}
\end{figure}
\begin{figure}
\iffigures
\setlength{\unitlength}{1cm}
\begin{picture}(18,8.7)
\put(0.7,0.7){\psfig{figure=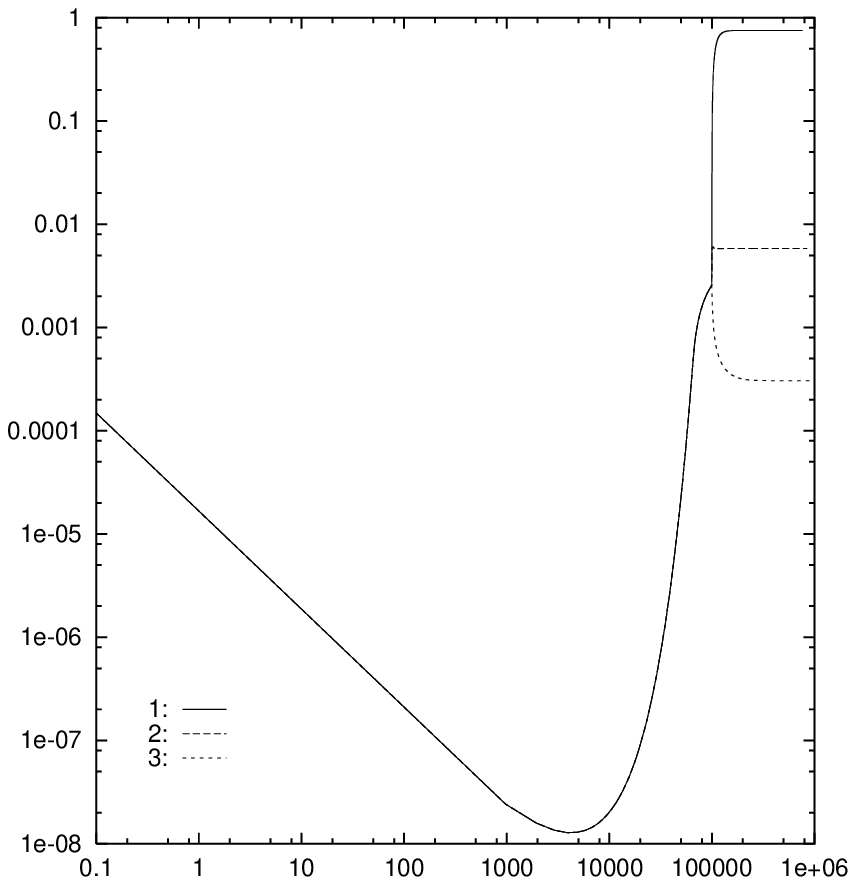,width=8 cm}}
\put(4.0,0.2){$ (t/t_{\rm 0})$}
\put(2.2,7.4){$1: t_{\rm acc}/t_{\ast}=0.1$}
\put(2.2,7.0){$2: t_{\rm acc}/t_{\ast}=1$}
\put(2.2,6.6){$3: t_{\rm acc}/t_{\ast}=10$}
\put(0.5,3.7){\rotate{$M_{\rm bh}/M_{\rm spheroid}$}}
\end{picture}
\else
\picplace{6.0cm}
\fi
\caption[]{\label{redif} The time evolution of the mass ratio of the central
black hole and its host spheroid with different correlation between starburst 
time scale and the turbulent accretion time scale. We calculate the cases: 
$t_{\rm acc}/t_{\ast} = 0.1, 1, 10$. We see that the variation of the final 
value $M_{\rm bh}/M_{\rm spheroid}$ is almost a factor of $10^{4}$ for a 
variation of time scale ratio by a factor of $10^{2}$. This shows that the 
correlation between these two time scales deeply dominates the final mass 
ratio limitation. It seems also that the mass ratio can finally converge to 
a limited region $\sim 0.006$ only when the starburst time scale and the 
turbulent accretion time scale are almost equal.}
\end{figure}

\section{Summary}

In this paper, we presented an $\it ab$ $\it initio $ model of
formation and evolution of early-type galaxies by combination of the
viscous accretion disk model, the star formation and nuclear activity
within the hierarchical galaxy formation scheme.

The basic ideas are:

1) An elliptical/bulge is formed by the merger of two disk galaxies in a
hierarchical universe.

2) Star formation and AGN evolution coexist in the disk evolution
stage (pre-merger) and also post-merger stage. They compete for the
gas supply and interact with each other.

3) The violent interaction between two disk galaxies by a merger can not
only form the spheroidal component of the elliptical or bulge, but
also drive the cool outer gas to the center, trigger a starburst
and a central AGN there. The merger triggered starburst probably can
stir up a turbulent accretion to feed both the starburst and AGN in
the center. These two activities interact with each other, feedback, 
thus speed up the BH growth, drain the gas in the molecular disk
quickly and constrain the final mass ratio $M_{\rm bh}/M_{\rm spheroid}
\sim 10^{-3}$.

From our calculation, we found:

1) Mergers can help to grow a massive BH in a very short time, and
shorten the convergence time for the mass ratio ($M_{\rm bh}/M_{\rm
spheroid}$) to a limited region.

2) The final mass ratio limitation does not depend on the mass of the
protogalaxy.

3) The exact time, when the merger happens after the disk evolution
quiets down, will not influence the mass ratio limitation.

4) Whether the ellipticals/bulges are formed by the major mergers or
minor mergers between two disk galaxies, or by multimergers is not
very critical for our final result.

5) The final mass ratio limitation does depend moderately on the inner
radius of the molecular disk $R_{\rm in}$, and the mass return rate
$R_e$ from star formation.

6) The exact number of the parameters $\beta_{1}, \beta_{2}$ for the
star formation and accretion time scale has no influence on the final
mass ratio limitation, but it is obvious that the correlation of these
two time scales (or $\beta_{1}, \beta_{2}$) is very critical to the
final mass ratio limitation.  It seems that the mass ratio ($M_{\rm
bh}/M_{\rm spheroid}$) can finally converge to a limited region near
0.006 only when the starburst time scale and the turbulent
accretion time scale can be approximately equal.

The conclusion 6) probably indicates a physical relationship between
starburst and central AGN as we discussed already in the previous
section. In the hierarchical universe, mergers will destroy the
quiescent disk galaxies, transfer the stars in the disks completely to
form a spheroidal component. At the same time, it will also drive a
large amount of gas into the central region and trigger a
starburst. The kinetic energy input to the ISM from the young massive
stars and supernovae can heat and shock the ISM, and probably induce
turbulent viscous accretion there. This turbulent viscous accretion
can feed both the starburst and central AGN, and grow a massive BH very 
quickly. The two activities, starburst
and central AGN can drain the gas in the disk in a very short time,
and also when the power of the central engine reaches some degree, it
probably can start a wind and help the starburst to blow out the hot
gas completely out of the disk, thus stop the accretion and star
formation there. 

In this paper, we did not include the kinetic energy term in our
viscosity directly, but from this simple model and calculation, it
seems that the feedback relationship between the starburst and central
engine can be a very interesting physical process. To do this, we
need more elaborated physical models and more numerical work.

\begin{acknowledgements} We should thank Dr. W. Duschl and
Prof. O. Gerhard very much for reading our manuscript carefully and giving
us a lot of helpful comments. We also should thank Dr. H. Falcke and
Dr. L. Wisotzki for many interesting discussions. YPW would
like to thank all the collegues in the group, C. Zier, T. Ensslin and
H. Seemann for their friendly help during YPW's stay in MPIFR.  PLB would like
to thank Dr. John Magorrian at Toronto for intense discussions of these
issues.  We also wish to thank the referee for helpful comments that helped
to improve the clarity of our arguments. 
YPW was first supported by a fellowship of the MPG, and then by the University
of Wuppertal; both sources of funding are gratefully acknowledged.

\end{acknowledgements}

\end{document}